\documentstyle[aps,epsfig,multicol]{revtex}

\begin{document}
\draft
\tighten
\newcommand{\onecolm}{\end{multicols}\vspace{-1.5\baselineskip}
\vspace{-\parskip}\noindent\rule{0.5\textwidth}{0.1ex}\rule{0.1ex}{2ex}
\hfill}
\newcommand{\twocolm}{\hfill\raisebox{-1.9ex}{\rule{0.1ex}{2ex}}
\rule{0.5\textwidth}{0.1ex}\vspace{-1.5\baselineskip} \vspace{-\parskip}
\begin{multicols} {2}}

\title{Macroscopic quantum fluctuations in noise-sustained optical patterns}
\author{Roberta Zambrini$^1$, Stephen M. Barnett$^{2}$,
Pere Colet$^1$ and Maxi San Miguel$^1$}
\address{$^1$ Instituto Mediterr\'aneo de Estudios Avanzados,
IMEDEA (CSIC-UIB),\\ Campus Universitat Illes Balears, E-07071
Palma de Mallorca, Spain.}
\address{$^2$Department of Physics and
Applied Physics, University of Strathclyde,\\ 107 Rottenrow,
Glasgow G4 0NG, Scotland }

\maketitle

\centerline{\today}

\begin{abstract}

We investigate quantum effects in pattern-formation for a degenerate
optical parametric oscillator with walk-off.  This device has a 
convective regime in which macroscopic patterns are both initiated and
sustained by quantum noise.  Familiar methods based on linearization
about a pseudo-classical field fail in this regime and new approaches
are required.  We employ a method in which the pump field is treated
as a $c$-number variable but is driven by the $c$-number representation
of the quantum sub-harmonic signal field.  This allows us to include the 
effects of the fluctuations in the signal on the pump, which in turn act
back on the signal. We find that the non-classical effects, in the form of squeezing,
survive just above the threshold of the convective regime. Further above threshold
the macroscopic quantum noise suppresses these effects.

\end{abstract}

\begin{multicols} {2}

\section{Introduction}
\label{introduction}

Nonlinear optics has provided an ideal testing ground for ideas in 
both nonlinear dynamics and quantum optics.  It provides fast 
nonlinearities and a degree of control that allow fundamental dynamical
systems to be realized and nonlinear phenomena like pattern formation to be
demonstrated \cite{ducci}.  It also provides systems with very low 
levels of noise so that fluctuations can be limited by quantum 
effects.  The combination of these features has led to the study of
quantum phenomena in optical pattern formation 
\cite{lugiato-castelli,patternsQuantum} and of 
noisy precursors of the patterns, which have been termed $quantum$ 
$images$ \cite{quantumimages,review,OPOale}.  
The accurate modeling of such quantum nonlinear systems
presents a significant challenge.  Pattern formation and dynamics is
usually associated with excitation of a large number of transverse
modes and a fully quantum description of each of these is required
in order to properly treat the quantum fluctuations. The Heisenberg 
picture produces a hierarchy of coupled nonlinear operator equations
that usually defies analysis. The preferred
method to date has been to linearize the quantum fluctuations about
a classical field amplitude that  usually takes a  constant value below 
threshold \cite{OPOale,jeffers}, but may be associated with a 
stable pattern above threshold
\cite{roby}.  

A more difficult situation arises if the system displays macroscopic
features driven by noise. In such cases we cannot expect linearization
of the quantum fluctuations to give reliable results and a new approach
is needed.  A simple device demonstrating macroscopic, noise-driven patterns
is the degenerate optical parametric oscillator in the presence of walk-off.
The semi-classical analysis of this device reveals a region of convective
instability, above the threshold for oscillation, in which noise sustained
structures are seen in the transverse field distribution \cite{marcoPRL,OPT_LETT}.
The aim of this paper is to develop a suitable approximation scheme with
which to model quantum effects in the parametric oscillator in this regime 
of operation.

The convectively unstable regime is characterized by an amplification and
flow of fluctuations \cite{deissler}. In systems in which the spatial reflection
invariance is broken by the presence of a group velocity term, local 
perturbations of the steady state can be advected more rapidly than their
growth rate. If the system is deterministic then at any fixed point any
initial localized perturbation decays and the system approaches the 
undisturbed steady state. In this case macroscopic patterns can arise and be
observed only if noise is continuously applied, the structure now being
regenerated at any time, hence the name noise-sustained patterns. These
structures are the result of noise self-organization, with magnification
factors of several order of magnitudes. They are thus interesting candidates
for the study of quantum correlations in spatially structured systems.

Any system with an advection (or drift or walk-off) term that is also not 
translationally invariant will, in general, be convectively unstable when 
operating sufficiently close to and above the onset of the instability of the
steady state. Hence, this type of instability has been predicted in a number of optical systems 
including Kerr media with a tilted pump \cite{marcoPRL,marco_symm} 
and optical parametric 
oscillators (OPO) with walk-off \cite{marco_symm,marcoPRE,gonzalo}.

Modeling quantum effects in the regime of convective instability for a 
nonlinear optical device presents a double challenge. First, the system 
has a broad spectrum both in frequency (at a fixed point) and in 
wave-vectors (far field at a fixed time), thus it cannot be studied within a 
few-mode approximation. Second, we should be able to follow the evolution  of
the fluctuations from the microscopic level through the amplification into  the
macroscopic pattern. In order to do this we introduce, in Sect.
\ref{steve.approx}, a suitable $non$-$linear$ approximation with which to treat
the convective regime of a degenerate optical  parametric oscillator.  In order
to fix the terms of reference for this approximation we begin, in Sect.
\ref{model}, with a review of the semi-classical features of the device and its
convective instability.  This is followed, in Sect. \ref{q.f.}, by a quantum
description of the device.  Once we have introduced our method we  discuss
quantum features of the device in its various regimes of operation (Sect.
\ref{trajectories}), paying particular attention to the demanding convective 
regime (Sect. \ref{convective}).


\section{ Semi-classical description and Convective regime}
\label{model}

We consider a Degenerate Optical Parametric Oscillator (DOPO), a
device consisting of a cavity filled with a $\chi^{(2)}$ nonlinear
medium, which converts a pump at frequency $2\omega$ into a
sub-harmonic signal at frequency $\omega$. The possibility of
phase matching the down-conversion process depends on the
birefringence of the crystal that provides a difference of
refractive index for differently polarized fields.  We can exploit
this difference in order to avoid the effects of dispersion by 
selecting the same index of refraction for the pump and signal
$n_{2\omega}=n_{\omega}$. In this paper we consider type-I phase 
matching for which ordinary polarized pump photons are down-converted 
to produce pairs of extraordinary polarized photons that are degenerate both 
in frequency and in polarization.

In anisotropic media rays do not necessarily travel in a direction
perpendicular to their wavefronts \cite{Saleh}.  
As a consequence the extraordinary-polarized signal generated in
our DOPO will walk off, that is it will propagate in the transverse
direction relative to the ordinarily-polarized pump. This transverse 
walk-off effect is described in the dynamical equations by a term which
accounts for a velocity relative to the frame of reference fixed
by a pump of finite transverse width.

The quantum effects we wish to study are associated with the convective 
regime and it is important to define this carefully.  The different 
regimes of operation of a DOPO can be understood within a semiclassical 
theory and this section provides a brief (semi-classical) analysis of the 
convective and other regimes.  A more complete discussion can be found in
\cite{marcoPRE}.  The intra-cavity field is described by two 
slowly varying complex field amplitudes $A_0(\vec x,t)$ and $A_1(\vec x,t)$ 
for the pump and signal respectively.  These depend on the transverse spatial
coordinates $\vec x = (x,y)$ and the time $t$. Within the paraxial
approximation (for propagation in the $z$ direction), the mean field limit 
and for single longitudinal mode
operation the dynamical equations become \cite{marcoPRE,oppo,ward}:
\begin{eqnarray}
\label{semiclpump}
\partial_t A_0(\vec x,t)&=& -\gamma_0[1+i\Delta_0-ia_0\nabla^2] A_0(\vec x,t)-
\frac{g}{2} A_1^2(\vec x,t) \nonumber \\
&&+E_0(\vec x)+\epsilon_0 \xi_0(\vec x,t),\\
\label{semiclsignal}
\partial_t A_1(\vec x,t)&=& -\gamma_1[1+i\Delta_1-ia_1\nabla^2-v\partial_y] 
A_1(\vec x,t) \nonumber \\
&&+g A_0(\vec x,t)A_1^*(\vec x,t)+ \epsilon_1\xi_1(\vec x,t).
\end{eqnarray}
Here $\xi_i~ (i=0,1)$
are additive Gaussian white sources of noise, with non-vanishing
correlations of the form:
\begin{eqnarray}\label{whitenoise}
<\xi_i(\vec x,t)\xi_j^*(\vec x',t')>=\delta_{ij}\delta(\vec x-\vec x')\delta(t-t').
\end{eqnarray}
The level of noise introduced is fixed by the parameters $\epsilon_0$ and 
$\epsilon_1$.  
Our fully quantum analysis will produce equations of similar form in which these
parameters are fixed.
$E_0$ is the amplitude of the driving field which we take to be real.
The remaining parameters in these equations are the cavity 
decay rates $\gamma_{i}$, the cavity detunings $\Delta_{i}$, the diffraction
$a_{i}$, the walk-off $v$ and the nonlinear coefficient $g$.
It is convenient to introduce scaled variables
\begin{eqnarray}
\label{scaling}
&&t'=\gamma t~,~~~~
\vec x'=\frac{\vec x}{\sqrt a}~,~~~~
v'=\frac{v}{\sqrt a}, \\
&&A_i'=\frac{g}{\gamma}A_i~,~~~~
E_0'=\frac{g}{\gamma^2}E_0~,~~~~
\epsilon_i'=\frac{g}{\gamma^{3/2}  a^{D/4}}\epsilon_i, \nonumber
\end{eqnarray}
where we have restricted the cavity decay rates and diffraction coefficients
such that $\gamma=\gamma_0=\gamma_1$ and $a=a_0=a_1/2$.  Our equations are 
valid either for one or two transverse spatial dimensions ($D=1,2$). 
On omitting the primes, our amplitude Eqs. (\ref{semiclpump})
and (\ref{semiclsignal}) become
\begin{eqnarray}
\label{semicl.eqpump}
\partial_t A_0(\vec x,t)&=& -[1+i\Delta_0-i\nabla^2] A_0(\vec x,t)-
\frac{1}{2} A_1^2(\vec x,t) \nonumber \\
&&+E_0(\vec x)+ \epsilon_0 \xi_0(\vec x,t),\\ 
\label{semicl.eqsignal}
\partial_t A_1(\vec x,t)&=& -[1+i\Delta_1-2i\nabla^2-v\partial_y] A_1(\vec x,t)
\nonumber \\
&&+A_0(\vec x,t)A_1^*(\vec x,t)+ \epsilon_1\xi_1(\vec x,t).
\end{eqnarray}
For a uniform driving field $E_0$, the equations (\ref{semicl.eqpump}) 
and (\ref{semicl.eqsignal}) admit the homogeneous stationary solution
\begin{equation}
\label{solst}
A_0^s=\frac{E_0}{1+i\Delta_0}~,~~~A_1^s=0~.
\end{equation}
The threshold for parametric oscillation can be determined by a
linear stability analysis of this solution. The linearized
equations for signal and pump fluctuations $\delta A_i(\vec
x,t)=A_i(\vec x,t)-A_i^s~ (i=0,1)$ are decoupled, and the
fluctuations of the pump are always damped. For the signal, we consider
perturbations of the form $e^{i\vec k\cdot\vec x+\lambda(\vec k)
t}$ and find the dispersion relation 
\begin{equation}\label{disp}
\lambda_\pm(\vec k)=-1+ivk_y\pm\sqrt{F^2-(\Delta_1+2|\vec k|^2)^2},
\end{equation}
where we have introduced a scaled pump
\begin{equation}
F=\frac{E_0}{\sqrt{1+\Delta_0^2}}.
\end{equation}
We find that there is an instability at $F=1$. For $F<1$, $Re(\lambda)<0$
and the solution (\ref{solst}) is absolutely stable. For $F>1$,
there is a positive growth rate of fluctuations ($Re(\lambda_+)>0$)
which takes a maximum value for $|\vec k_c|=\sqrt{-\Delta_1/2}$ if
the signal detuning is negative ($\Delta_1<0$), and for $k=0$ if
$\Delta_1>0$. In this paper we are interested in the case of
pattern formation and we restrict our analysis to the case
$\Delta_1<0$. The instability at $F=1$ when $v=0$ is a Turing
instability, in which a stationary pattern appears \cite{oppo}. If
$v\neq 0$ then the eigenvalue becomes complex and we find a Hopf
bifurcation in which a traveling pattern emerges \cite{marcoPRE}.

The direction of instability is determined by the eigenfunctions
$V_\pm(\vec k,-\vec k)$ of the linear problem $\partial_t
V_\pm(\vec k,-\vec k)= \lambda_\pm(\vec k) V_\pm(\vec k,-\vec k)$.
Solving this gives
\begin{eqnarray}
\label{inst_dir}
&&V_\pm(\vec k,-\vec k)=e^{i\Phi_\pm}\delta A_1(\vec k)\pm\delta A_1^*(-\vec k)\\
&&e^{i\Phi_\pm(\vec k)}= 
\mp\frac{i\Delta_1+2i |\vec k|^2\mp \sqrt{|A_0^s|^2-(\Delta_1+2 |\vec k|^2)^2}}
{A_0^s}.
\nonumber
\end{eqnarray}
The solution $V_+(\vec k,-\vec k)$ gives the direction of amplification of
fluctuations, while fluctuations are damped for $V_-(\vec k,-\vec
k)$. In particular, for the critical wave-vector $|\vec k_c|$ and
for a real pump $E_0$ and $\Delta_0=0$, we obtain  $V_\pm(\vec
k_c,-\vec k_c)=\delta A_1(\vec k_c)\pm \delta A_1^*(-\vec k_c)$.
Therefore, in this case, the difference of real parts and the sum
of imaginary parts of field in $\vec k_c$ and $-\vec k_c$ will
show damped fluctuations at threshold. We also note that the
instability direction is independent of the walk-off term.

\begin{figure}
\epsfysize=4.5cm \centerline{\epsfbox{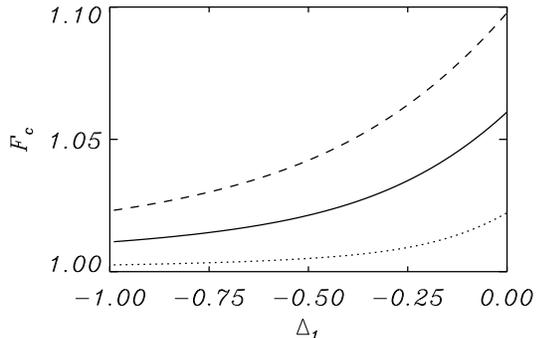}} \caption[]{Stability
diagram as a function of the signal detuning $\Delta_1$.  The
different lines correspond to  the threshold
of absolute instability ($F_c$) for different values of the  walk-off
parameter: 
$v=0.2$ (dotted line), $v=0.42$ (continuous line),
$v=0.6$ (dashed line). When $F<1$ the solution (\ref{solst}) is
absolutely stable, while for $1<F<F_c$ the solution is
convectively unstable. } \label{fig1}
\end{figure}

Above the instability threshold (F=1) the steady state (\ref{solst}) is
convectively unstable: any perturbation grows while travelling in the
direction fixed by the walk-off term and eventually leaves the system
\cite{marcoPRE}. 
In this regime a continuous perturbation such as a source of
noise, gives rise to a noise-sustained pattern consisting in disordered
travelling stripes in the signal. On increasing the pump a second threshold
is reached at $F=F_c$. Beyond this threshold the pattern is sustained by
the nonlinear dynamics, being also present in the absence of
perturbations, once it is formed. The state (\ref{solst}) 
is  absolutely unstable in this regime \cite{marcoPRE}. 
In Fig. \ref{fig1} we plot the result of the calculation of the absolute
instability threshold as function of the signal detuning
$\Delta_1$, for different values of the walk-off parameter $v$.

Walk-off has three main effects in this process
of pattern formation \cite{marco_symm,marcoPRE}. The first is the
existence of the convective regime in which patterns are sustained
by noise. Second is that it breaks the rotational symmetry,
favoring the formation of stripes orthogonal to the walk-off
direction and traveling in this direction. Thirdly the selected
wave-vector, that is the most intense mode $\vec k_M$ of the pattern,
depends on the walk-off parameter. 
An approximate expression for $\vec k_M$ can be obtained in the
context of front propagation into an unstable state 
\cite{OPT_LETT,marcomajid}.

There are two important characteristics of the noise sustained
patterns that exist in the convective regime. 
The first is a broad spectrum, both in frequency and in wave-vectors
\cite{spectrum}.
 Second is the presence of $macroscopic$ amplified signal
fluctuations around the unstable reference state (\ref{solst}).
These characteristics imply that the convective regime cannot be
studied within a few-mode approximation, because many modes
contribute significantly  to the spectral properties. The presence of
macroscopic fluctuations also invalidates approximations based on
linearization schemes. These facts make a quantum formulation of
the convective regime especially difficult. We face a situation in
which nonlinearities determine the dynamics of fluctuations around
the reference state, with fundamental quantum noise being
amplified by several orders of magnitude to produce a macroscopic
pattern in the signal.


\section{Quantum formulation}
\label{q.f.}

In the quantum formulation of the DOPO  the intracavity pump and
signal fields are given by operators $\hat A_0(\vec x,t)$ and
$\hat A_1(\vec x,t)$ that satisfy standard equal-time commutation
relations \cite{OPOale}
\begin{equation}
\left[\hat A_i(\vec x, t), \hat A_j^\dagger(\vec x', t)\right]=\delta_{ij}\delta(\vec x-
\vec x') \, ,
\label{equaltimecommutator}
\end{equation}where the indices $i,j$ stand for $0,1$.
Following the techniques described in \cite{OPOale}, we can introduce a model
Hamiltonian for the device.  This will include the effects of diffraction
together with the driving by a real, classical external field, nonlinear
interaction between the fields and cavity damping.  Our model, however, 
also requires that we take account of the effects of walk-off.  The 
resulting Hamiltonian gives, on making the usual Markov approximation, 
the coupled Heisenberg equations:
\begin{eqnarray}
\label{heis.eq}
\partial_t{\hat A_0} (\vec x,t) &=& -
 \gamma_0[(1+i\Delta_0)-ia_0\nabla^2]{\hat A_0}(\vec x,t) \nonumber \\
 &&-\frac{g}{2}\hat A_1^2(\vec x,t)+E_0(\vec x)+ \hat F_0\\
 \partial_t{\hat A_1} (\vec x,t) &=& -
\gamma_1 [(1+i\Delta_1)-ia_1\nabla^2-v \partial_y]{\hat A_1}(\vec x,t)\nonumber \\
 &&+g\hat A_0(\vec x,t)\hat A_1^\dagger(\vec x,t)+ \hat F_1
\end{eqnarray}
Note that these are very similar in form to the semi-classical Eqs.
(\ref{semicl.eqpump}) and (\ref{semicl.eqsignal}).  
The Langevin operators $\hat F_i$ describe the quantum
noise added as a consequence of the interaction with the bath of external
modes.  These have the non-vanishing second moments:
\begin{equation}
\langle \hat {F_i} (\vec x,t)\hat {F_j}^\dag  (\vec x',t')\rangle
= 2\gamma_i\delta_{ij}\delta(\vec x-\vec x')\delta(t-t').
\end{equation}

A direct solution of these non-linear Langevin equations of
operators is impractical, requiring the solution of an infinite
hierarchy of equations for the evolution of all the products of
operators that are coupled by the dynamics. A standard alternative
approach to this Heisenberg picture is to consider the evolution
equation of the reduced density operator $\hat\rho$ of the system in
the Schr\"odinger picture and to use quasi-probability
functionals. In this approach to the quantum dynamics of open
problems, the intracavity dynamics is described by a master
equation \cite{steve}:
\begin{equation}\label{master}
\frac{\partial\hat\rho}{\partial t}=\frac{1}{i\hbar}[\hat H, \hat \rho]+
\Lambda \hat \rho ,
\end{equation}
where $\hat H$ is the Hamiltonian
\onecolm
\begin{eqnarray}\nonumber
\hat H=\hbar \int d^2 \vec x
\sum_{j=0,1}
\left[\gamma_j \hat {A}_j^\dag(\vec
x)(\Delta_j-a_j\nabla^2)\hat {A}_j(\vec x)\right]+
i\gamma_1v \hat {A}_1^\dag(\vec x) \partial_y\hat {A}_1(\vec x) +
i E_0(\vec x)
(\hat {A}_0^\dag(\vec x)-\hat {A}_0(\vec x))+
i\frac{g}{2}(\hat {A}_0(\vec x)\hat A_1^{\dag^2}(\vec x)-
h.c.).
\end{eqnarray}
\twocolm
\noindent
The Liouvillian $\Lambda$ accounts for dissipation through the
partially reflecting mirrors of the cavity and is given by \cite{OPOale}
\begin{eqnarray}
\Lambda\hat \rho = \sum_{j=0, 1}\int d^2\vec x \gamma_j\left\{[\hat A_j(\vec x), 
\hat \rho
\hat A_j^\dagger(\vec x)]+[\hat A_j(\vec x)\hat \rho, \hat A_j^\dagger(\vec x)] \right\}
\nonumber.
\end{eqnarray}
The master equation  (\ref{master}) can be mapped onto an equation of 
motion for one of a number of quasi-probability distributions in the 
phase-space of the system \cite{steve,gardiner,carmichael}.  These 
distributions are functionals of the $c$-number fields $\alpha_i(\vec x)$ associated
with the operators $\hat A_i(\vec x)$. This evolution equation is obtained
by substituting products of field operators
and the density operator, depending on the ordering,
by suitable operations on the distribution functionals
\cite{OPOale,roby}.

The evolution equations obtained in this way for the distributions
are functional partial differential equations.  These are not in general of 
the Fokker-Planck  type and do not lead to well-behaved stochastic representations
in terms of Langevin equations driven by Gaussian white noise. 
In particular the Hamiltonian term describing the $\chi^2$ interaction gives a contribution:
\begin{eqnarray}
[\hat {A}_0(\vec x)\hat A_1^{\dag^2}(\vec x)-
h.c.,\hat \rho]&&\Longleftrightarrow 
\left(s\alpha_0\frac{\delta^2}{\delta \alpha_{1}^2}+
\frac{1-s^2}{4}\frac{\delta^3}{\delta \alpha_{1}^2\delta \alpha_{0}^*}\right.\nonumber\\
&&\left.+\frac{\delta}{\delta \alpha_{0}}\alpha_1^2-
2\alpha_{0}\alpha_1^*\frac{\delta}{\delta \alpha_{1}}+c.c.\right)W_s.\nonumber
\end{eqnarray}
where the parameter $s$ depends on the ordering.
This term does not fulfill the requirements that guarantee
a positive definite solution  for $W_s$: in the Wigner representation ($s=0$)
we find third order derivatives, while it is known \cite{pawula}
that positiveness requires a Fokker-Planck form of the master 
equation (only first and second order derivatives) or to include derivatives 
to all orders.
For the $P$ ($s=1$) and $Q$ ($s=-1$) representation third order derivatives disappear, but the
diffusion matrix is not positive definite so that positive solutions are again not
guaranteed, although the $Q$ retains positivity through having a minimum allowed 
width \cite{gardiner}. Generally these
problems have been avoided by using linearization schemes \cite{carmichael}. 
Such linearization approximations, however, are valid only for small 
damped fluctuations.  They cannot be used in a convective regime as the
reference state is unstable and the fluctuations, far from being 
small, are amplified.  The alternative of the $P$ positive representation 
\cite{P+} is not suitable for the same reason and the unstable reference 
state results in diverging trajectories.

These problems of the convective regime can be illustrated by
comparison with the situation of a DOPO below the threshold of
signal generation. In this case the stable solution is a
homogeneous pump with an amplitude that depends on the coherent
driving field. The signal field is zero on average, but its
fluctuations show a level of self-organization that increases near
the threshold. This is the regime of {\it quantum images}
\cite{quantumimages,review,OPOale}, 
noisy precursors generated by quantum noise.
These images reflect the presence of eigenmodes of the linearized
equations, whose eigenvalues are such that their negative real
part approaches zero at threshold.  The fluctuations of these
eigenmodes are the least damped ones and dominate the dynamics of
the signal. The important point is that the intensity of such
quantum images of the signal is of the order of quantum noise,
while the pump has a macroscopic mean value.  It is then possible
to neglect the fluctuations in the pump, approximating it by a
classical coherent field \cite{carmichael}. In this approximation
the Hamiltonian is a quadratic function of the quantum operators.
The consequence is that a  well defined Fokker-Planck equation for
the Wigner distribution is obtained. Such Fokker-Planck equations
can be represented in terms of stochastic Langevin
equations for the $c$-number field $\alpha_1(\vec x)$ \cite{OPOale}.
The same type of approximation, linearizing around a pattern
solution \cite{roby}, is generally possible in the absolutely
unstable regime  above threshold. A common feature of these two
regimes (absolutely stable and unstable) is that quantum noise does
not change drastically the solution with respect to the stable
deterministic solution.  This means that in the stochastic
representation, fluctuations only induce the trajectory to visit a
small region in phase space in the neighborhood of the
deterministic solution.  In the convective regime the classical
deterministic solution is unstable and {\it macroscopically} different
from the stochastic solution. In this regime quantum noise in the
DOPO is amplified, destroying the zero-valued homogeneous deterministic
solution for the down-converted field and driving the system into
noise sustained states having a macroscopic number of photons.


\section{Time dependent parametric approximation}
\label{steve.approx}

In this section we we propose an approximate description of the
quantum dynamics of the DOPO in the convective regime, based on
the main physical features of this regime.
  Our aim is to be able to treat the macroscopic
quantum fluctuations associated with the signal field in the 
convective regime.

In the convective regime there are large signal fluctuations
around the unstable solution $A_1=0$. The coupling of signal and
pump gives the nonlinear saturation for these amplified
fluctuations. On the other hand, the pump field is always
macroscopic and stable, with small damped fluctuations. This
suggests the approximation of neglecting quantum noise in the pump
and approximating it by a classical field ${\mathcal A}_0(\vec
x,t)$. In this way we obtain a Hamiltonian that is quadratic in the
operators describing the quantum dynamics of the signal
field. For such quadratic Hamiltonians, the Wigner
quasi-probability functional of the complex function
$\alpha_1(\vec x,t)$ obeys the following Fokker-Planck equation in
which the classical pump field ${\mathcal A}_0(\vec x,t)$ appears
parametrically:
\onecolm
\begin{eqnarray}
\nonumber
\frac{\partial W(\alpha_1;{\mathcal A}_0)}{\partial t}=&&
\left[-\left(\frac{\delta}{\delta\alpha_1}\gamma_1 [(1+i\Delta_1)-ia_1\nabla^2 
- v{\partial_y}]  \alpha_1(\vec x,t)+g{\mathcal A}_0(\vec x,t)
\alpha_1^*(\vec x,t)    +c.c.\right) 
+ \gamma_1 \frac{\delta^2}{\delta\alpha_1\delta\alpha_1^*}
 \right] W(\alpha_1;{\mathcal A}_0)).
\end{eqnarray}
\twocolm
The associated Langevin equation that represents the stochastic
dynamics of the signal field $\alpha_1(\vec x,t)$ is 
\begin{eqnarray}
\label{LA1}
 \partial_t \alpha_1(\vec x,t)&=& -
  \gamma_1 [(1+i\Delta_1)-ia_1\nabla^2 - v {\partial_y}]
  \alpha_1(\vec x,t)+\nonumber\\
&&g{\mathcal A}_0(\vec x,t)
\alpha_1^*(\vec x,t)+\sqrt {\gamma_1}\xi_1(\vec x,t),
\end{eqnarray}
where $\xi_1(\vec x,t)$ is a complex Gaussian white noise (see
Eq.(\ref{whitenoise})).
This noise term accurately represents the effects of vacuum 
fluctuations associated with cavity losses on the signal field.
We note that treating the pump field classically in this way is
a natural extension of the parametric approximation to three-mode
interactions, which treats a strong mode classically and has been 
widely used in quantum optics for many years \cite{Louisell}.

It is important to note that ${\mathcal A}_0(\vec x,t)$ cannot be
replaced by an expectation value of $\langle\hat A_0\rangle$ as
would be possible in the regime of absolute stability (quantum
images). Such an ansatz decorrelates the pump modes from the
sub-harmonic ones and eliminates the saturation effect of the
pump. In fact, with such an ansatz Eq. (\ref{LA1}) becomes linear,
giving a Gaussian probability distribution for the signal modes.
This distribution would always be centered on zero, but with 
statistical moments that diverge above threshold
because the signal modes are undamped in the convective
regime. Therefore, the stochastic differential equation must be
solved self-consistently with an equation defining the dynamics of
the classical field ${\mathcal A}_0$. The equation we propose 
for ${\mathcal A}_0$  is suggested by the
Heisenberg equation (\ref{heis.eq}), with $\hat
A_0$ replaced by a classical field ${\mathcal A}_0$. We first neglect the
noise source in (\ref{heis.eq}) since quantum fluctuations
entering in the cavity are unimportant, as compared with the
macroscopic fluctuations of the signal term $\hat A_1^2$. 
Secondly  we replace the operator $\hat A_1^2$ by
the $c$-number function $\alpha_1^2$ associated with our stochastic
representation of the signal.  This replacement is independent 
of operator ordering and hence will be the same should be choose
to use a different quasi-probability.  This procedure gives a 
partial differential equation for the ``classical'' pump field
driven by the $c$-number representation of the {\it quantum}
signal field:
\begin{eqnarray}
\partial_t{\mathcal A}_0 (\vec x,t) &=& -\gamma_0[(1+i\Delta_0)-ia_0\nabla^2]
{{\mathcal A}_0}(\vec x,t)\nonumber\\ \label{classicalpump}
&& -\frac{g}{2}\alpha_1^2(\vec x,t)+E_0(\vec x)
\end{eqnarray}
A justification for this equation is that its mean value coincides
with the expectation value for $\hat A_0$ obtained from the
operator equation (\ref{heis.eq}).  This procedure is reminiscent of
the time-dependent refinement of the parametric approximation described 
in some detail by Kumar and Mehta \cite{Kumar}.  This approach allows
for the quantum evolution of the weak fields to feed back and affect 
the classical strong field.  In the approach of Kumar and Mehta, this
feedback is via quan\-tum expectation values of opera\-tors for the weak
fields.  Here, however, we are required to take explicit account of the
noisy properties of the quantum sub-harmonic field.  We do this by 
using the $c$-number representation of the quantum field, associated
with our stochastic simulation of it, as a term in equation (\ref{classicalpump}).

In summary, our time dependent parametric approximation is defined
by stochastic classical equations in the Wigner representation for
the fields ${\mathcal A}_0$ and  $\alpha_1$, which, with the scaling
(\ref{scaling}), are:
\begin{eqnarray}
 \nonumber
\partial_t{\mathcal A}_0 (\vec x,t) &=& -[(1+i\Delta_0)-i\nabla^2]{{\mathcal A}_0}(\vec x,t)
\\ &&-\frac{1}{2}\alpha_1^2(\vec x,t)+E_0(\vec x)\label{langevinpump}
\\ \nonumber
\partial_t \alpha_1(\vec x,t)&=& -
\left[(1+i\Delta_1)-2i\nabla^2 - v {\partial_y}\right]\alpha_1(\vec x,t)\\
\label{langevinsignal}&&+{\mathcal A}_0(\vec x,t)
\alpha_1^*(\vec x,t)+\frac{1}{a^{D/4}}\frac{g}{\gamma}\xi_1(\vec x,t).
\end{eqnarray}
Stochastic averages of the $c$-number variable $\alpha_1(\vec x,t)$
will provide symmetrically ordered averages of the quan\-tum fluctuations
in the signal field as driven by the ``classical'' pump field.
The classical pump field is driven by the macroscopic quan\-tum
fluctuations in the signal as represented by the $c$-number
representation $\alpha_1(\vec x,t)$.
This time dependent parametric approximation appears useful in
situations in which there are large fluctuations of the signal
that cannot be described by approximations based on linearization. 


\section{Stochastic trajectories and Wigner distribution function}
\label{trajectories}

Numerical simulation of the stochastic trajectories associated
with the Langevin equations (\ref{langevinpump}) and 
(\ref{langevinsignal}) gives a good intuitive
understanding of the dynamical properties of the regime below
threshold, the convective regime and the absolutely unstable
regime. In this section we present such numerical simulations 
working with a single transverse dimension $(D=1)$ \cite{numerical}. \\
\indent Fig. \ref{evolutNF} 
is a space-time plot of the near-field for the signal in the 
below-threshold, convective and absolutely unstable regimes.
Fig. \ref{evolutFF} gives the far-fields associated with the
same simulations.  In the following we discuss the different properties 
of these trajectories and how they are reflected in the associated Wigner
distribution. In particular, we consider the phase space dynamics
of the most intense modes of the signal pattern. The Wigner
probability distribution associated with these modes displays distinctly 
non-Gaussian features in the convective regime.  These are a result 
of the interplay of non-linear and walk-off effects.  It is clear that
they cannot be described within a linearization scheme that does
not take account of this interplay between the signal and pump
fields.
\onecolm
\begin{figure}
\centerline{\epsfysize=6.5cm \epsfbox{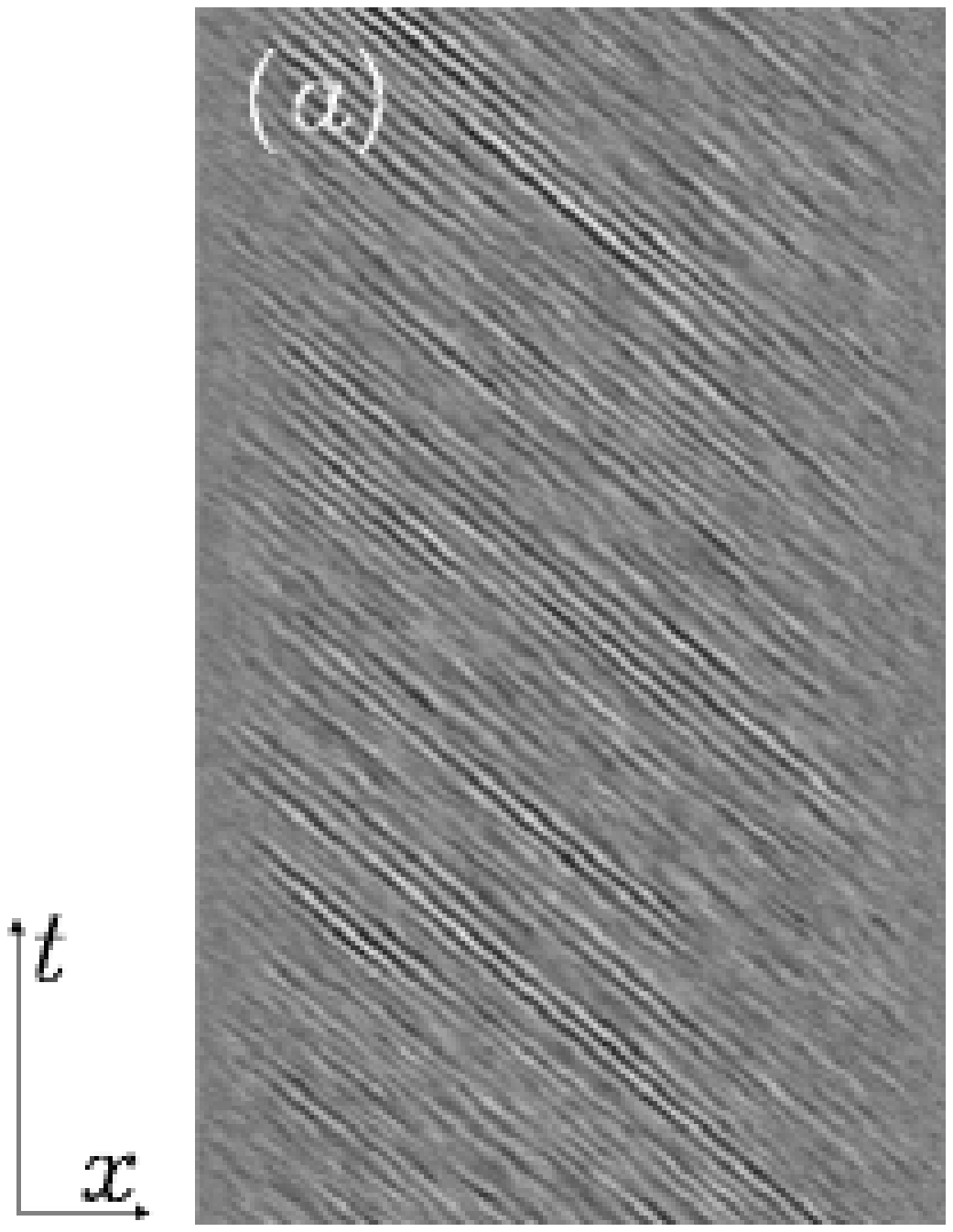}\hspace{1cm}
\epsfysize=6.5cm \epsfbox{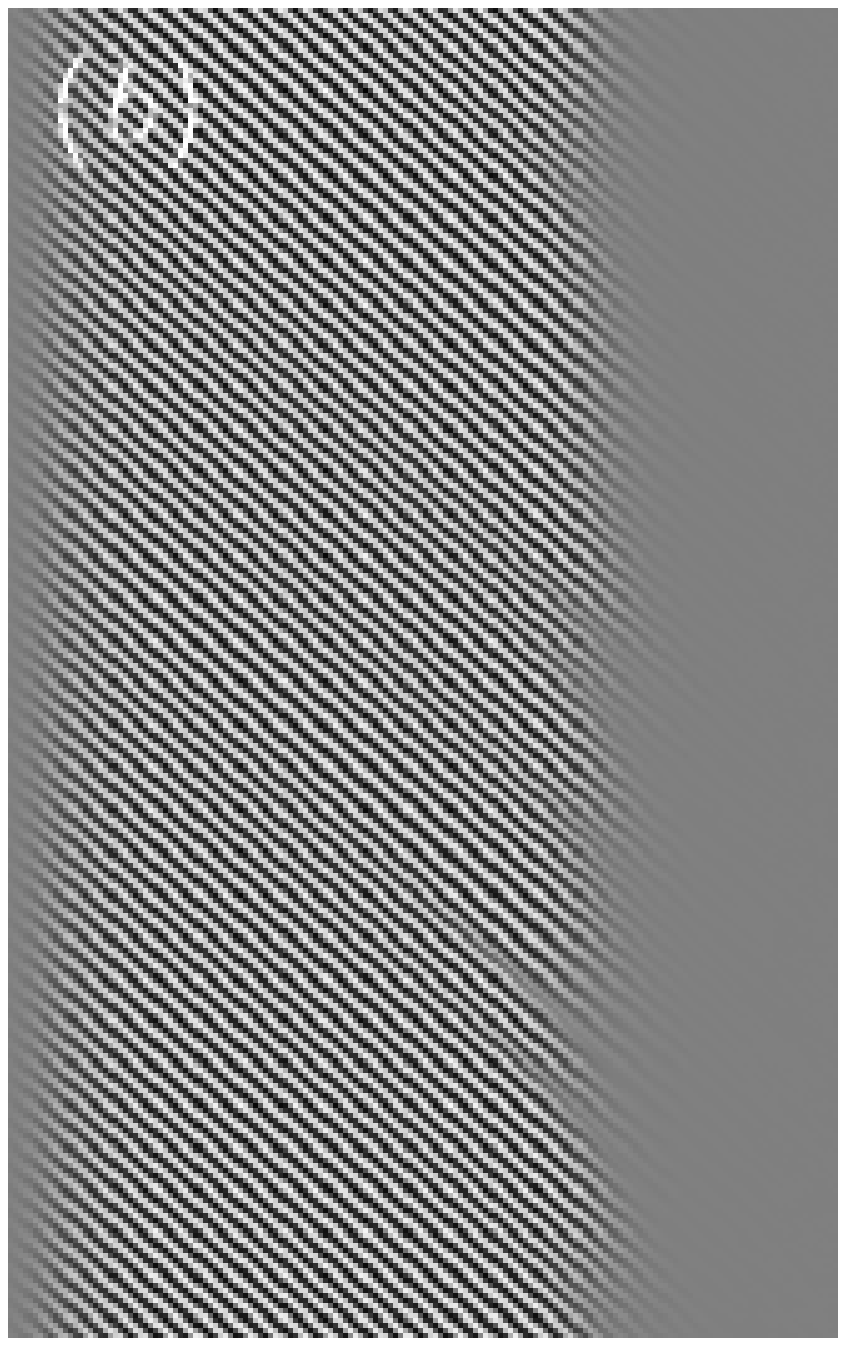}\hspace{1cm} \epsfysize=6.5cm
\epsfbox{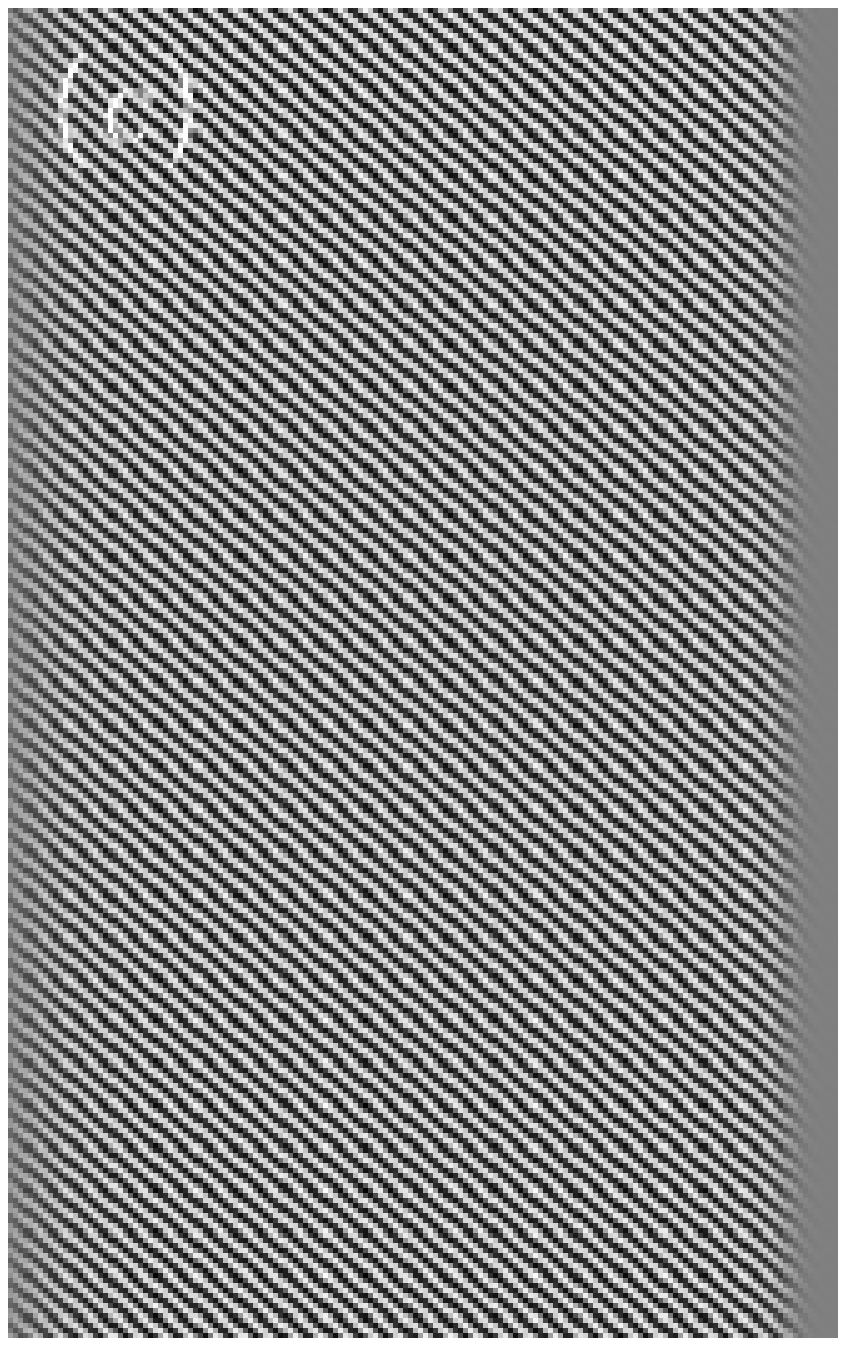}\vspace{0.2cm}} \caption[]{ Evolution of the near field of the
real part of the signal $Re(\alpha_1(x,t))$ for: ($a$)
$F=0.999$, ($b$) $F=1.025$, ($c$) $F=1.1$, in $2500$
time units. Parameters are: $\Delta_0=0,\Delta_1=-0.25,v=0.42$,
system size $=1.7678\cdot512$ pixels  $\simeq 900$ space units.
Only the regions in which the signal is excited are shown, that is
the region of the plateau of the supergaussian pump
\cite{supergaussian}.} \label{evolutNF}
\end{figure}

\begin{figure}
\centerline{\epsfysize=6cm \epsfbox{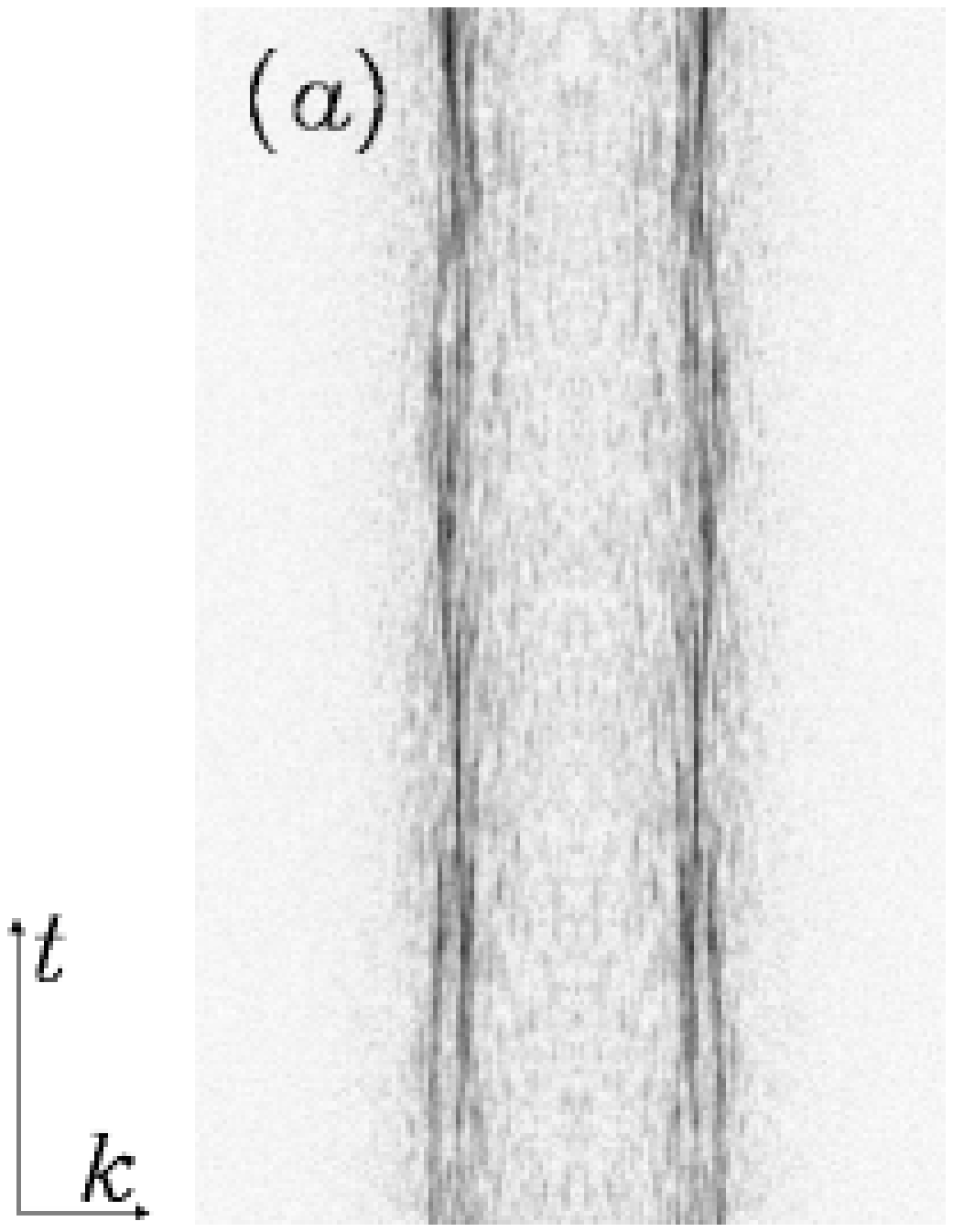} \hspace{1cm}
\epsfysize=6cm \epsfbox{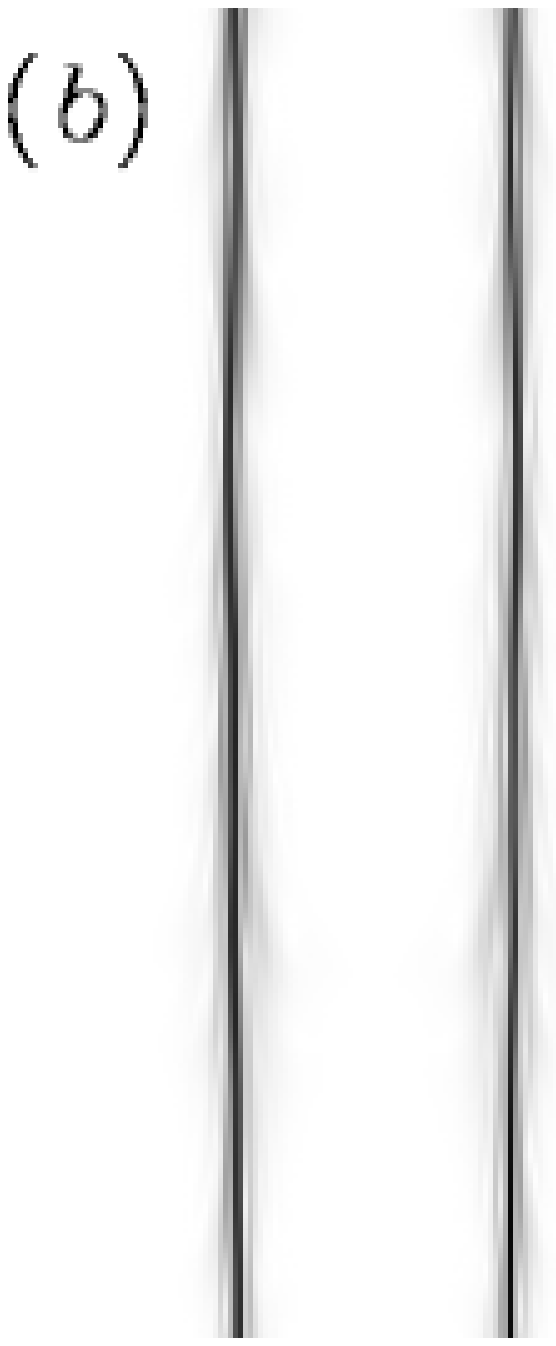}\hspace{1cm} \epsfysize=6cm
\epsfbox{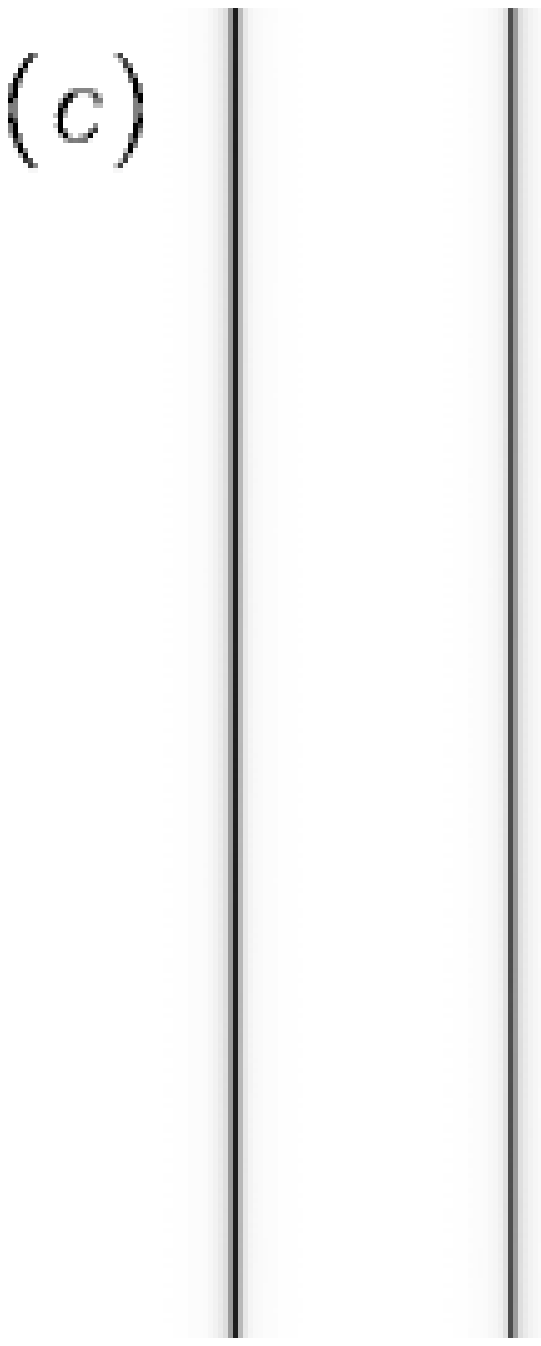}} \caption[]{Evolution of the far field  $|\alpha_1(k,t)|$ for: ($a$)
$F=0.999$, ($b$) $F=1.025$, ($c$) $F=1.1$. Same
parameters as in Fig. \ref{evolutNF}} \label{evolutFF}
\end{figure}
\twocolm

\subsection{Below threshold}

Below but close to threshold we find weakly damped  fluctuations which
are a precursor to the traveling pattern that
appears at threshold.  The fluctuations impose a degree of spatial
self-organization in those regions in which the pump is sufficiently 
strong to bring the OPO close to threshold. In Fig. \ref{evolutNF}($a$)
we plot the real part of the stochastic variable $\alpha_1(x,t)$ for a single
trajectory. This is a realization of these fluctuations for a pump with a
supergaussian profile \cite{supergaussian}.  Noisy patterns of this form 
have been predicted for the below-threshold OPO without walk-off
and have been termed quantum images \cite{quantumimages,review,OPOale}. 
Not too close to threshold, the damped
fluctuations  can be analyzed with
linearization procedures \cite{OPOale} in the limit of small
fluctuations. Our nonlinear
quantum equations enable us to study also the regime closer to
threshold, where large critical fluctuations are expected to occur. 
Note,
in particular, that the results in Fig. \ref{evolutNF}($a$)
were obtained for $F=0.999$.

The selection of a preferred wave-number in the stochastic pattern
of Fig. \ref{evolutNF}($a$) becomes more evident in the far
field shown in Fig. \ref{evolutFF}($a$). It is clear that there
are preferred values of the wave-number but that  a broad distribution
of weakly damped modes around these preferred $k_c$ and $-k_c$ modes
is apparent.
\begin{figure}
\centerline{\hspace{-0.1cm}\epsfysize=4cm \epsfbox{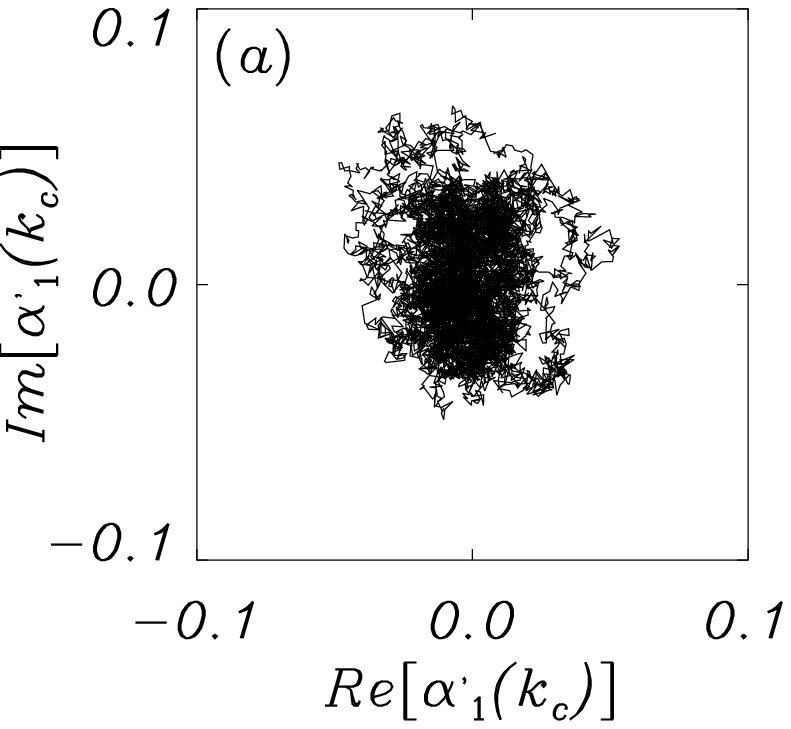} 
\hspace{0.2cm}\epsfysize=4cm\epsfbox{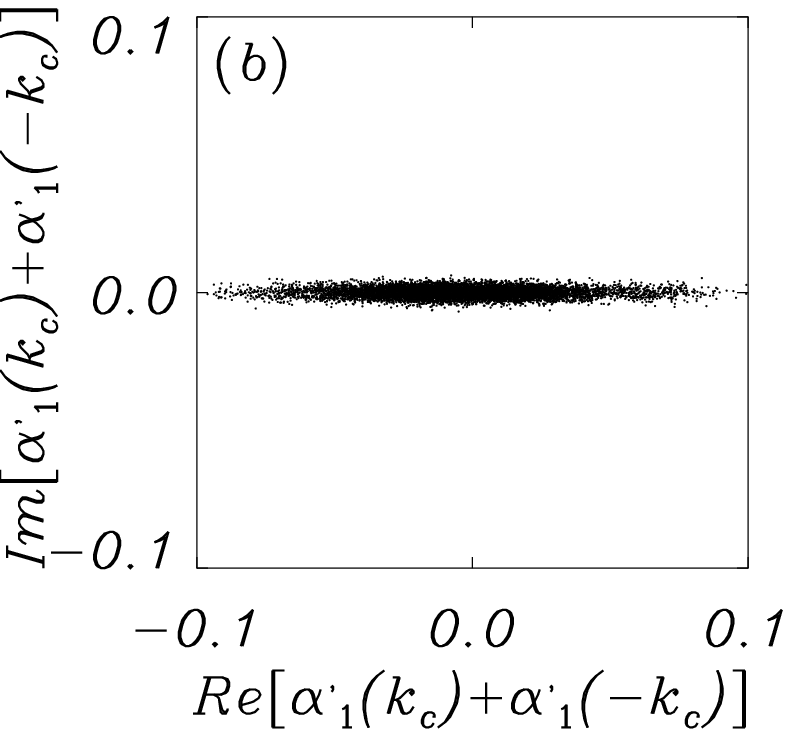}} 
\caption[]{($a$) trajectory of the slowly
varying amplitude $\alpha'_1(+k_c)$ during 20.000 time units.
($b$) trajectory of $[\alpha'_1(+k_c)+\alpha'_1(-k_c)]$ during
20.000 time units. $F=0.999$, other parameters are as in
Fig. \ref{evolutNF}, except for $dx=51\lambda_c/512\simeq 1.7702$,
where $\lambda_c=2\pi/k_c$, 512 is the number of grid points. }
\label{tra.below}
\end{figure}
An interesting characterization of the stochastic dynamics in the
far field, Fig. \ref{evolutFF}($a$), is obtained by looking at
the time evolution of the stochastic amplitudes for the most intense 
modes $\alpha_1(k_c,t)$.  We first recall that the linear stability 
analysis of Sect. \ref{model} identifies the existence of a non-vanishing 
frequency ($\omega(k)=vk$) at threshold caused by the walk-off. This 
implies that a traveling pattern will emerge above threshold and that 
the corresponding Fourier modes will oscillate at this frequency. We can
remove this time-dependence by working in a frame rotating at this 
frequency.  This corresponds to factoring out a time factor $e^{i\omega(k) t}$
to obtaining the slowly varying amplitude 
$\alpha'_1(k,t)=\alpha_1(k,t)e^{-i\omega(k) t}$.
 A phase space trajectory for the slowly varying amplitude of the
dominating Fourier component, $\alpha'_1(k_c,t)$, is shown in 
Fig. \ref{tra.below}($a$). 

The linear stability analysis of Sect. 
\ref{model} also identified the direction of instability $V_+$. 
In particular, in the case of a real pump, and for the critical mode 
$k_c$, this direction is given by $[\alpha_1(+k_c)+\alpha^*_1(-k_c)]$. 
As a consequence the superposition of modes $[\alpha_1(+k_c)+\alpha_1(-k_c)]$ 
can be decomposed in two quadratures, one corresponding to the
direction of instability that becomes undamped at threshold
($Re[\alpha_1(+k_c)+\alpha_1(-k_c)]$), and the orthogonal quadrature
($Im[\alpha_1(+k_c)+\alpha_1(-k_c)]$)  that remains
damped.
We observe that the superposition of slowly varying 
modes $\alpha_1'(\pm k_c,t)$ can be decomposed into damped and undamped
quadrature in the same way.
In fact due to the symmetry  $\omega(k)=-\omega(-k)$ we have
$V_\pm(\vec k,-\vec k)=$ $e^{i\omega(k)t}
[e^{i\Phi_\pm}\delta A'_1(\vec k)\pm\delta A_1'^*(-\vec k)]$,
so that the relative phase $e^{i\Phi_\pm}$ between the slowly varying modes
is the same as that in the equation (\ref{inst_dir}).
Hence, we can also identify the real and imaginary quadratures of 
the superposition of modes $[\alpha_1'(+k_c)+\alpha_1'(-k_c)]$
as damped and
undamped at threshold. 
The corresponding time trajectory of this superposition 
of modes displays very clearly the expected reduction of fluctuations 
in the damped imaginary quadrature (see Fig. \ref{tra.below}($b$)). 

\begin{figure}
\epsfig{file=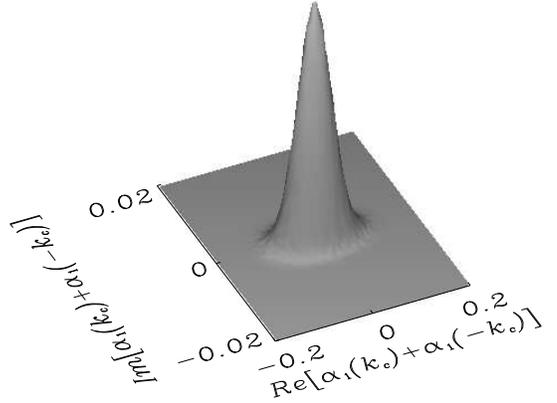,clip=true,width=0.45\textwidth}
\caption[]{ Wigner distribution for the superposition of modes
$\alpha_1(+k_c)+\alpha_1(-k_c)$. Parameters of Fig.\ref{tra.below}.
Total time 2.000.000 units. Note the factor 10 difference in the scale 
of the two axes.}
\label{prob.below}
\end{figure}
From the stochastic trajectories that randomly visit the different
points of phase space it is easy to construct a relative histogram
giving a probability density in this phase space. This density 
is identified with the Wigner distribution.  As with all Wigner functions,
the marginal distributions, obtained for one field quadrature by integrating
over the orthogonal quadrature, are true probability distributions for
the remaining quadrature. 
At a finite distance from threshold the Wigner distribution
$W(\alpha_1(k))$ for the field $\alpha_1(k)$ obtained in this way has
a Gaussian shape consistent with a linearized analysis of
fluctuations. Such a Gaussian Wigner distribution is a  solution of the
Fokker-Planck equation for the Wigner representation of {\it
linear} signal fluctuations. If we consider the Wigner
distribution for the superposition of modes discussed above
$W(\alpha_1(+k_c)+\alpha_1(-k_c))$, then we obtain a Gaussian centered
on the origin but with a variance that depends on the orientation
in phase space \cite{symm}. There is an axis with a reduced variance
(`squeezed') and the orthogonal one with a larger variance
(`anti-squeezed') (see Fig. \ref{prob.below}). These features reflect 
the asymmetry or phase-sensitivity of the fluctuations already visualized 
in the stochastic trajectory. 


\subsection{Convective regime}
\label{conv.traj}

Differences between the regime below threshold and the convective
regime are clearly seen both in the near and far signal fields. We
observe a macroscopic traveling pattern in the near field (Fig. \ref{evolutNF}($b$)).  This is clearly associated with wave-numbers distributed around 
the value of the selected one ($k_M$) in the far field (Fig. \ref{evolutFF}($b$)).
The spectrum of excited wave-numbers is clearly  narrower in the convective
regime than below threshold.  This is reflected in the more regular pattern
appearing in the near field.
Our simulations display the typical features associated with the convective 
regime \cite{marcoPRE}:
\begin{itemize}
\item The noise sustained pattern does not fill the whole region in which 
the pump has a value above threshold.  This is because
the pattern grows while traveling in the walk-off direction. Note
that the space point at which the pattern reaches a macroscopic
observable value changes randomly from time to time.  This reflects the
origin of the pattern in (quantum) noise.
\item The far field shows the predominance of different wave-numbers at different
times resulting in a spatial spectrum that is broader than that found in the absence 
of walk-off or in the absolutely unstable regime. There is competition between the 
modes within this broad spectrum and hence it is not possible to
define, in this regime, a {\it single} wave-number $k_M$ corresponding to the
most excited modes. Modes with different wave-numbers compete to form the 
pattern, switching on and off as the pattern evolves.
\end{itemize}

Phase space trajectories for this regime are shown in Fig.
\ref{tra.conv}.  We find that there are random changes in the 
phase and amplitude of the slowly varying signal $\alpha'_1(+k_c)$
around a zero mean value (Fig. \ref{tra.conv}($a$)). This is similar to the behavior depicted in
Fig. \ref{tra.below}($a$)  below threshold.
The difference is that in the convective regime macroscopic intensities 
are reached, with the signal amplitude taking values comparable to
those reached in the absolutely unstable regime
(compare scales of Figs. \ref{tra.below}($a$), \ref{tra.conv}($a$),
and \ref{anelli}). The continuous changes in intensity from zero
to macroscopic values originate in the fact that, in the convective 
regime, a given mode is not constantly switched-on (see
Fig.\ref{evolutFF}($b$)). The pattern is sustained by noise
and is subject to a continuous renovation: different stripe patterns
(with different wave-numbers) grow, travel in the system starting
from noise and die out. This has an important consequence in the
time scales of the far field dynamics: below threshold these
scales are determined by noise, while in the convective regime
they are determined by the time needed for a perturbation to travel
through the system. Another indication of the nonlinear dynamics
of fluctuations that occur in this regime is that the quadrature
displaying reduced fluctuations is no longer the one determined by the
linear analysis. This is seen in Fig. \ref{tra.conv}($b$)
where the ellipse of fluctuations is tilted with respect to the
corresponding one below threshold Fig. \ref{tra.below}($b$).
\begin{figure}
\centerline{\hspace{-1cm}\epsfysize=4cm \epsfbox{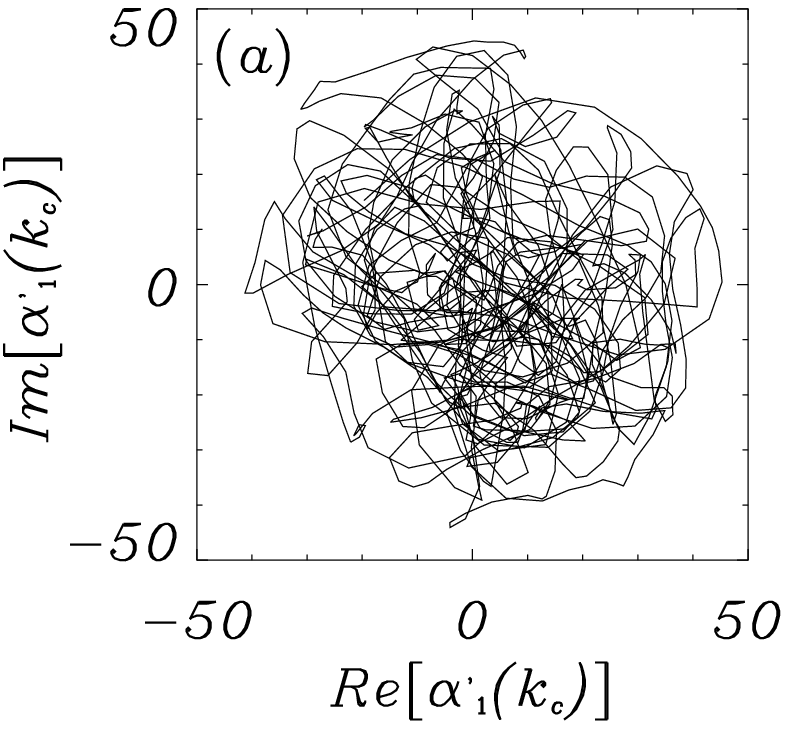} 
\hspace{0.2cm}\epsfysize=4cm\epsfbox{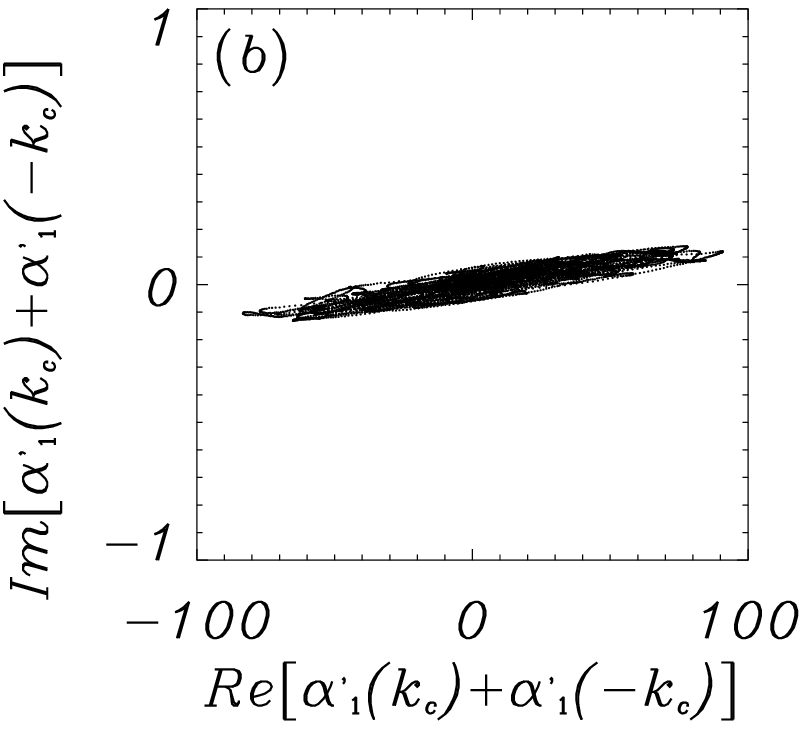}} 
 \caption[]{ ($a$) Trajectory of slowly
varying amplitude of $\alpha'_1(+k_c)$ during 100.000 time units.
($b$) Trajectory of $[\alpha'_1(+k_c)+\alpha'_1(-k_c)]$.
Parameters $F=1.025$,
$\Delta_0=0,\Delta_1=-0.25,v=0.42,dx\simeq 1.7702$. }
\label{tra.conv}
\end{figure}
\begin{figure}
\epsfig{file=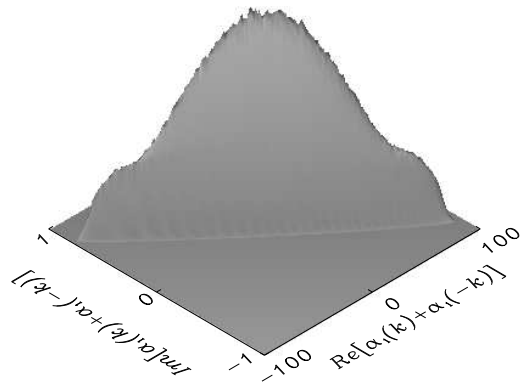,clip=true,width=0.4\textwidth} \caption[]{
W($\alpha_1(+k)+\alpha_1(-k)$), for an
excited mode $k=1.04k_c$, obtained from a trajectory during
10.000.000 time units. Other parameters as in Fig.
\ref{tra.conv}.} \label{prob.conv}
\end{figure}

The  probability distributions obtained from the trajectories of
Fig. \ref{tra.conv} also reflect the nonlinear nature of the
fluctuations in this regime. In Fig. \ref{prob.conv} we show the
$W$ distribution for the superposition of modes
$[\alpha_1(+k)+\alpha_1(-k)]$ for one of the most excited
wave-numbers, namely $k=1.04k_c$. A most noticeable feature is
the non-Gaussian shape of the distribution for large values of the
amplitude in the direction of undamped fluctuations. The wings
of the distribution originate in the macroscopic fluctuations of
the mode under consideration when it switches-on. Its most
probable value is still zero, reflecting the fact that most of the time the
mode remains switched-off.
We can view these nongaussian features in the wings of
our Wigner functions as precursors of the pair of peaks appearing 
in the absolutely unstable regime. 
These wings become more pronounced as we approach the absolutely unstable
regime.

Finally, we note that the modes that become excited and contribute
to the dynamics seem to reach a common maximum amplitude.  This is 
probably fixed by the maximum value of the energy exchanged with 
the pump mode in the nonlinear interaction. This is shown in 
Fig.\ref{3_campane} where the possible values of different modes are 
seen to be cut-off at essentially the same amplitude.  The nongaussian
form of these distributions is also clear and this again demonstrates
that we are dealing with nonlinear effects associated with the quantum
fluctuations.  
\begin{figure}
\hspace{-1cm}
\centerline{\epsfysize=5.5cm
\epsfbox{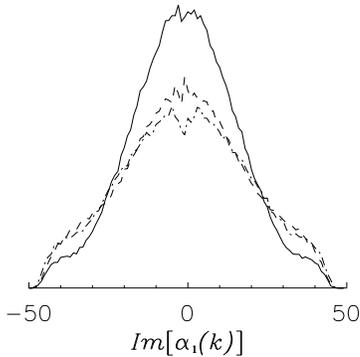}}
\caption[]{Section of the Wigner distribution along the imaginary axis 
$W(0,Im(\alpha_1(k)))$ for 3 excited modes:
$k_c$ (dashed line), $k'=1.04k_c$  (dash dot line), $k''=1.06k_c$
(continuous line). Same parameters as Fig. \ref{tra.conv}.}
\label{3_campane}
\end{figure}
\subsection{Absolutely unstable regime}
In the absolutely unstable regime we observe from the near
field plot, Fig. \ref{evolutNF}($c$), that a macroscopic and stable
traveling pattern fills the whole of the above threshold region.
This behavior is reflected in the far field, Fig. \ref{evolutFF}($c$),
which shows a well-defined and fixed dominant wave number and a narrow 
spatial spectrum.
We should note that the dominant wave-number $k_M$
does not coincide with the most unstable wave-number at threshold
($k_c$). This is a consequence of the interplay between
nonlinearities and walk-off. Phase
space trajectories for the amplitudes of these two modes are shown
in Fig. \ref{anelli}. Even after elimination of the rapid
frequency there remains a phase diffusion process, but
macroscopic values of the intensity are maintained. Although there is
essentially only  the phase diffusion for  $k_M$, the critical mode,
with wave-number $k_c$, displays a second frequency superimposed on 
the phase diffusion process.
\begin{figure}
\centerline{\hspace{-0.1cm}\epsfysize=4cm \epsfbox{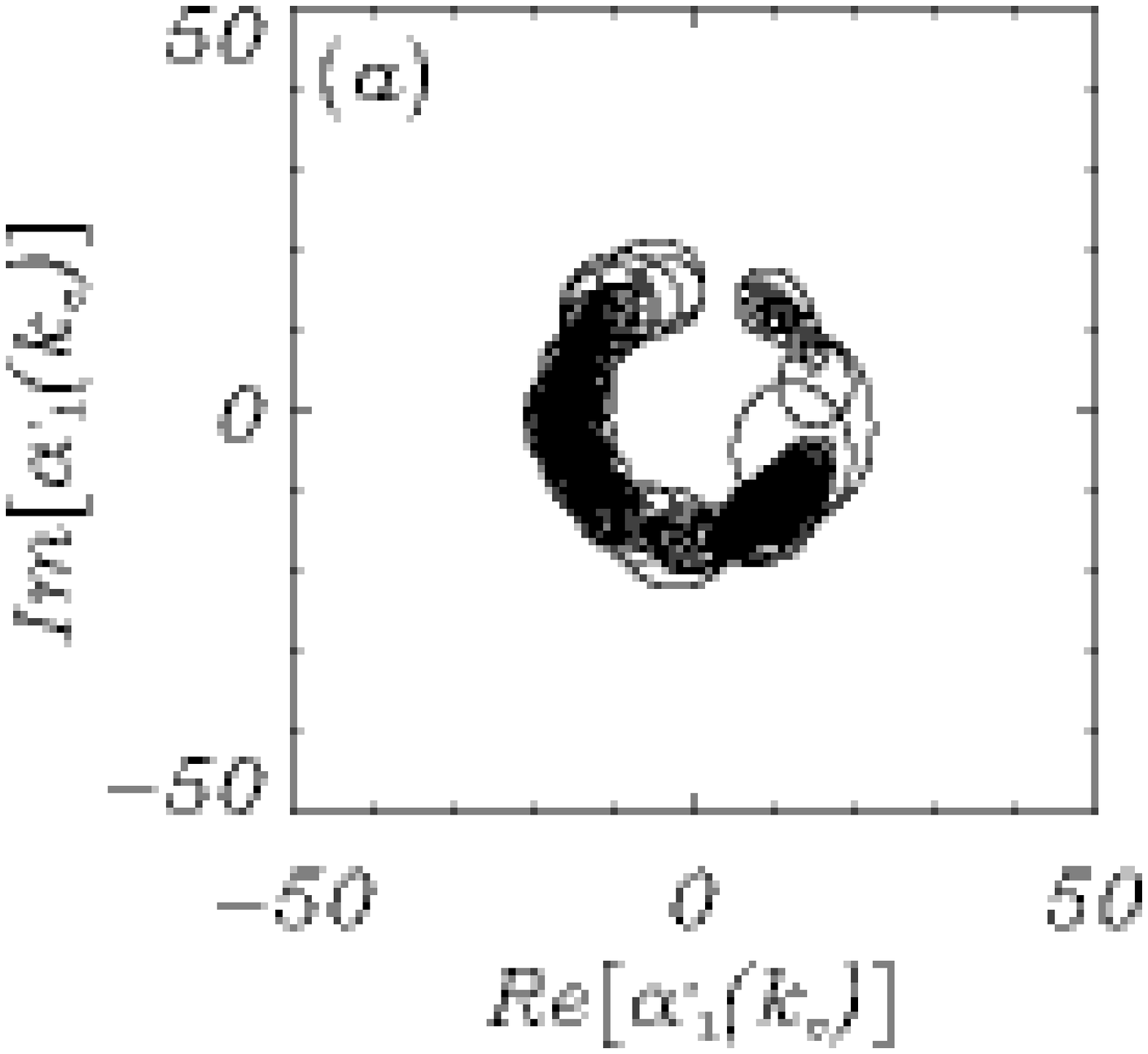} 
\hspace{0.2cm}\epsfysize=4cm\epsfbox{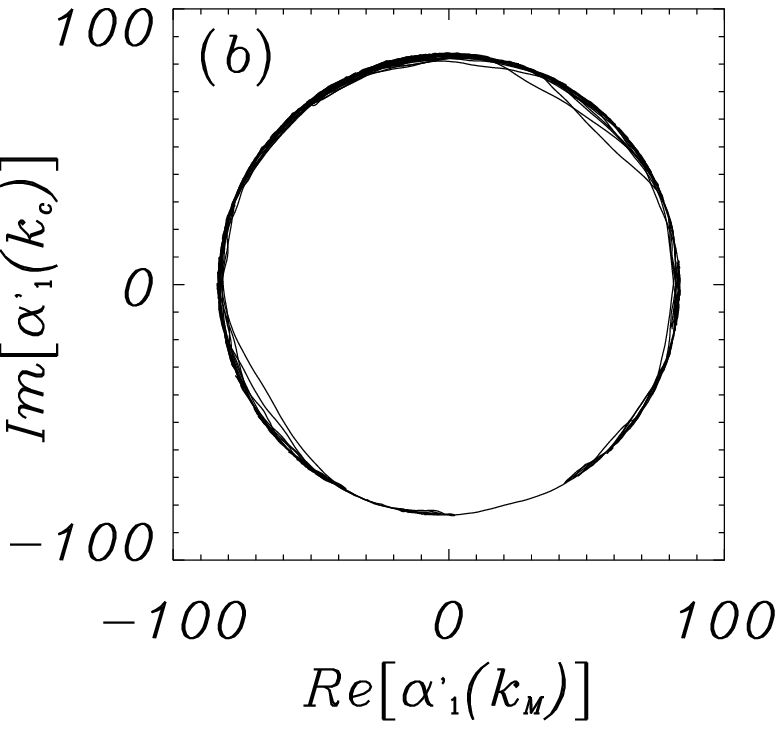}} 
\caption[]{Trajectories of ($a$) $\alpha'_1(k_c)$
and ($b$) $\alpha'_1(k_M)$  during 100.000 time units,
$F=1.05$, other parameters as in Fig. \ref{tra.conv}}
\label{anelli}
\end{figure}
\begin{figure}
\centerline{ \epsfysize=3.5cm \epsfbox{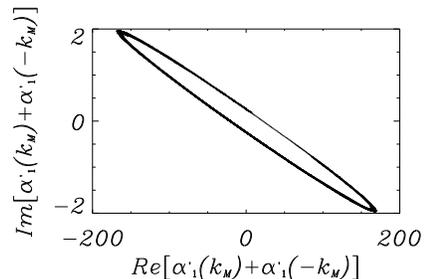}}
\caption[]{Trajectory of $[\alpha'_1(+k_M)+\alpha'_1(-k_M)]$ over
10.000.000 time units. Other parameters as in Fig.\ref{anelli}. }
\label{ellisse}
\end{figure}

The phase space trajectory for the superposition of modes
$[\alpha'_1(\vec k_M,t)+\alpha'_1(-\vec k_M,t)]$ is shown in Fig.
\ref{ellisse}.  We observe that fluctuations are not uniformly
distributed around a zero value as they were in the below threshold
(Fig. \ref{tra.below}) and convective (Fig. \ref{tra.conv})
regimes.  Instead, they describe a closed curve around the
origin. The associated $W$ distributions display peaks at two values.
These correspond to the two points of maximum curvature of the 
elliptical ring.
\begin{figure}
\centerline{\epsfysize=5.cm \epsfbox{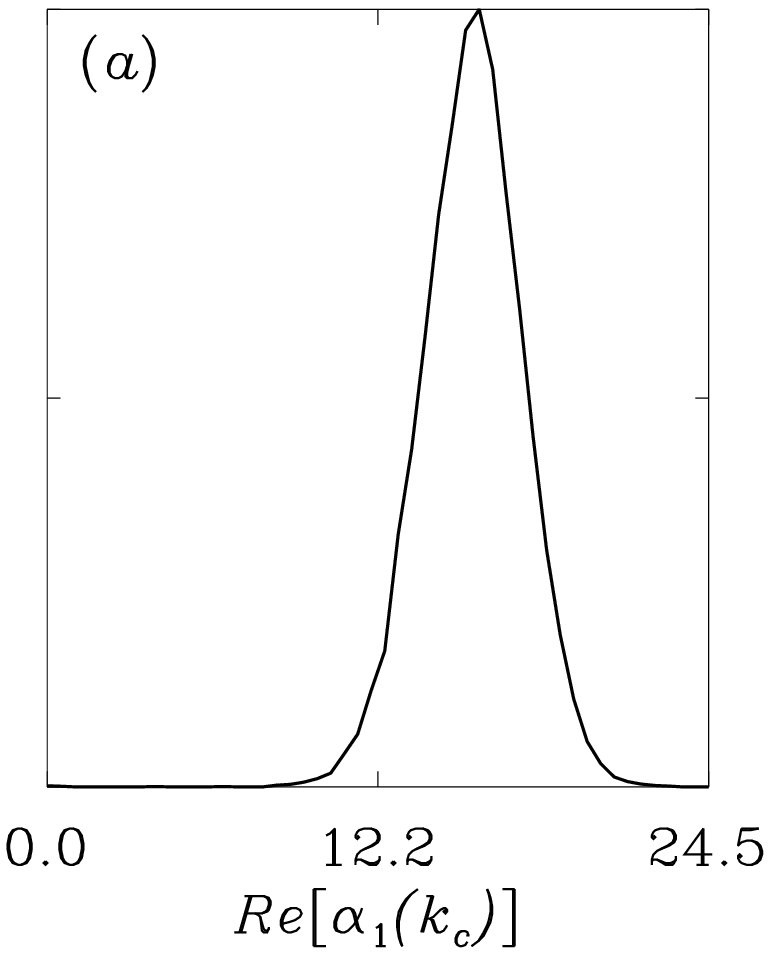}
\epsfysize=5.cm \epsfbox{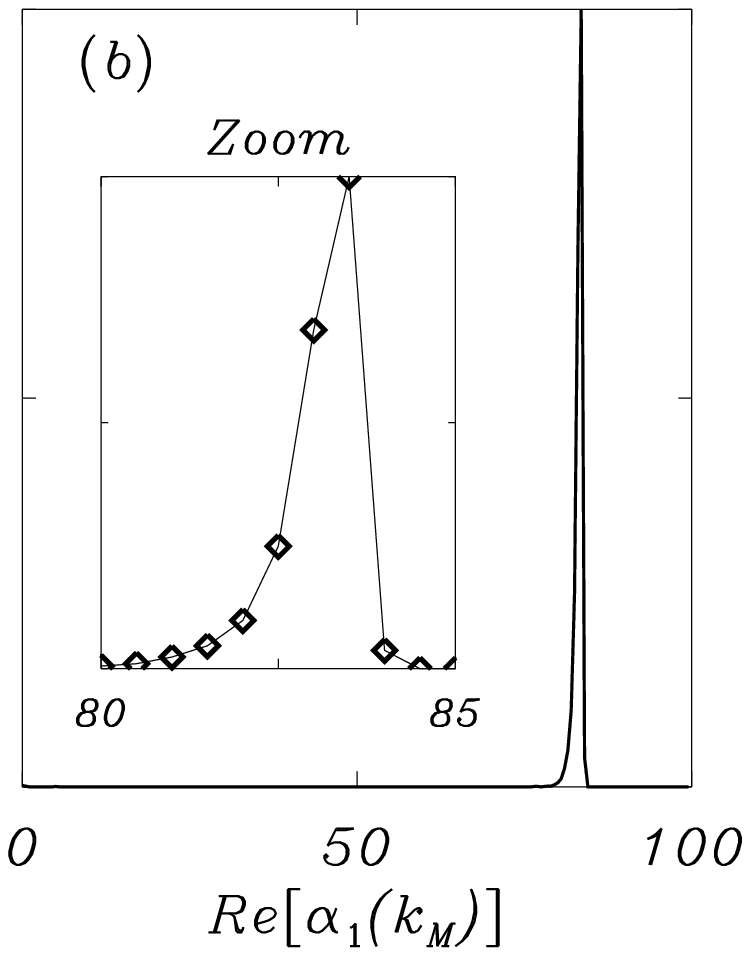}}
\caption[]{$W(Re(\alpha_1(k)),0)$, for positive values
of $Re[\alpha_1(k)]$, for ($a$): $k=k_c$, ($b$): $k=k_M$, obtained
from a trajectory during  10.000.000 time units.
Other parameters as in Fig. \ref{anelli}.
These figures are symmetric around $0$.}
\label{cut.abs}\end{figure}

The main characteristics of the trajectories in phase space are
reflected in the associated Wigner distributions. For the less intense 
modes contributing to the dynamics we can approximate the associated 
Wigner function $W(\alpha_1(k))$ by a Gaussian, displaced from and 
orbiting about the origin in phase space. In Fig.\ref{cut.abs}($a$) we 
show a cut along the real direction 
of the Wigner distribution for the critical 
mode ($W(Re[\alpha_1(k_c)],0)$).
By contrast, the most intense mode (with wave-number $k_M$) displays 
some interesting new features. Fig.\ref{cut.abs}($b$) shows 
an asymmetry in the distribution of fluctuations around the mean 
amplitude in each of the peaks, with a sharp decay of the distribution at some
maximum amplitude. These facts indicate the existence of nonlinear 
properties associated with the quantum fluctuations in the absolutely unstable 
regime. These features would necessarily be absent in any analysis based on a 
linearization about a deterministic macroscopic state.


\section{Non-classical properties in the convective regime.}
\label{convective}

The convective regime is characterized by amplified fluctuations
and macroscopic noisy patterns.  It is interesting to ask, therefore,
if any of the low-noise quantum features found below threshold
can survive in this noisy environment.  Quantum effects in the 
OPO have been observed as sub-shot noise fluctuations both in the 
field quadratures and intensity differences associated with the
down-converted light \cite{squeez}.  Examples of the noisy features
associated with the real part of the signal field in this regime 
are plotted in Fig. \ref{signals} for three different values of the 
driving field, all within the convective regime.  Note the different
scales on the vertical axes in these figures.
\begin{figure}
\centerline{\epsfysize=8cm \epsfbox{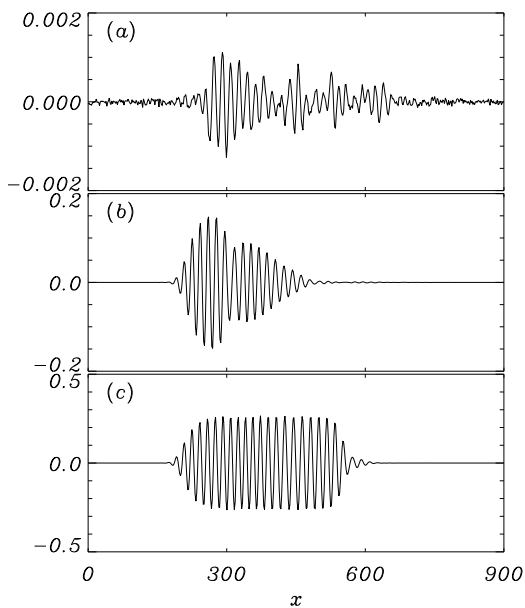}} \caption{ Snapshots
of the real part signal $Re(\alpha_1(x))$   for different pump
values:$(a)$ $F=1.001$, $(b)$  $F=1.01$, $(c)$  $F=1.025$. Other
parameters are: $dx=1.7678$, 512 grid points, $a_0=1$,
$\Delta_0=0$, $\Delta_1=-0.25$, $v=0.42$.  Note the different 
vertical scales in the figures.} \label{signals}
\end{figure}

It is helpful, in looking for non-classical effects, to keep in
mind the manner in which such effects appear below threshold.
We will also restrict ourselves to the study of quantum correlations 
in the far field.  Conditions for squeezing and associated non-classical
effects are usually expressed in terms of normally ordered moments
of operators (indicated by $:$ $:$).  These can be obtained from
the symmetrically ordered moments (indicated by $S()$), that are
associated with the Wigner function, by use of the commutation
relations (\ref{equaltimecommutator}):
\begin{eqnarray}
\nonumber
\langle: \hat{A}(k,t)\hat{A}(k',t):\rangle&=&
\langle S(\hat{A}(k,t)\hat{A}(k',t))\rangle\\ \nonumber
\langle: \hat{A}(k,t)\hat{A}^{\dagger}(k',t):\rangle&=&
\langle S(\hat{A}(k,t)\hat{A}^{\dagger}(k',t))\rangle-\frac{1}{2}\delta(k-k')
\end{eqnarray}
The $\delta$ function appearing in the second of these equations
is a signature of the shot or vacuum noise.  Our approximation scheme 
is based in the Wigner representation and gives results for correlations 
of symmetrically ordered operators for the intracavity fields.  
In order to obtain results for the 
corresponding normally ordered products and to test for the presence of
non-classical effects, we need to establish a reference shot 
noise level. 
This level can be obtained for each quadrature correlation from the
variance of the $linear$ stochastic process associated with the empty
cavity:
\begin{eqnarray}
\nonumber
\partial_t s(x,t)&=& -
\left[(1+i\Delta_1)-2i\nabla^2 \right]s(x,t)
+\frac{1}{{a}^{1/4}}\frac{g}{\gamma}\xi_1(x,t).
\end{eqnarray}
Here we have omitted the walk-off term as it does not affect the shot 
noise level.  Squeezing in our simulations will be associated with a
quadrature probability distribution that is narrower than the Gaussian
associated with this linear process.  In general we can consider a 
different quadrature for each wave-number $k$.  It is useful to define
a pair of (superposition mode) quadratures for each $k$ parameterized 
by the angle $\theta$.
For the critical wave-number these take the form
\begin{equation}
\label{quaddefn}
\hat{X}_\pm(\theta)=\frac{1}{2}\left[\hat{A}_1(k_c,t)\pm
\hat{A}_1(-k_c,t)\right]e^{i\theta}+h.c..
\end{equation}
We expect, in general, that the most strongly squeezed quadrature
 should depend on the value
of $\theta$ \cite{jeffers}.

We begin our investigation of the convective regime at a point that
is just above threshold with
$F=1.001$ (Fig. \ref{signals}($a$)). Fluctuations 
associated with the pattern are in this case  still relatively 
small and we find that the Wigner distribution has a Gaussian shape 
as shown in Fig.\ref{gauss1001}.  We find that there is quadrature
squeezing, with the squeezed quadrature 
$\hat{X}_-(0)$ exhibiting
the same level of squeezing as is found just below threshold.
In particular, for $F = F_{thr}\pm 0.001$ we find that the intracavity
field is squeezed by $50\%$ below the shot noise limit for a flat pump 
and by $37\%$ for a supergaussian pump \cite{output}. 
This indicates a smooth variation across threshold for the squeezed
quadrature variance.
For excited modes, other
than the critical one, we also find squeezing below the shot noise level
for the appropriate quadrature.
\begin{figure}\vspace{-0.8cm}
\centerline{\epsfysize=5.5cm \epsfbox{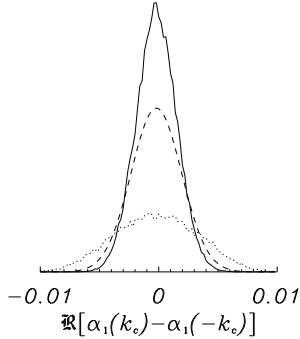}} 
\caption[]{$W(Re(\alpha_1(k_c,t)-\alpha_1(- k_c,t)),0)$.
  Continuous line is obtained for $F=1.001$, 
and dotted line for $F=1.01$. The dashed line represents the distribution 
for the vacuum state, corresponding to the shot noise level. The distributions are
relative to  trajectories of 2.000.000 time units.
Other  parameters as in Fig. \ref{signals}.} 
\label{gauss1001}
\end{figure}

Increasing the value of the pump, so as to move further into
the convective regime, leads to a rapid increase in the magnitude 
of the signal field.  Indeed, for ($F=1.01$) we observe,
in Fig. \ref{signals}($b$) that the signal field has grown
by two orders of magnitude.  Fluctuations are still extremely
phase sensitive and, as depicted in Fig. \ref{tra.conv}($a$) 
there is a strong reduction in the fluctuations for some quadratures.
This reduction is insufficient, however, to reach below the shot noise
level and there is no squeezing.  In fact, we find residual fluctuations 
$+27\%$ $above$ the shot noise level.  This is comparable with the 
value associated with the coherent states.  These enhanced fluctuations 
are associated with a much broader Wigner distribution as shown in 
Fig. \ref{gauss1001}.  It is remarkable, however, that this enhanced 
but still small level of fluctuation can coexist with the macroscopic
fluctuations in orthogonal quadrature.  If we move still further 
above threshold then we find, for $F=1.025$
(Fig. \ref{signals}($c$)), a variance which is $159$ times the
shot noise level and both quadratures display fluctuations that are 
well above the level usually associated with quantum effects. 
We note that for the parameter values used here, the threshold of 
absolute instability for an infinite system occurs at $F\sim 1.035$.

\begin{figure}
\centerline{\epsfysize=3.cm \epsfbox{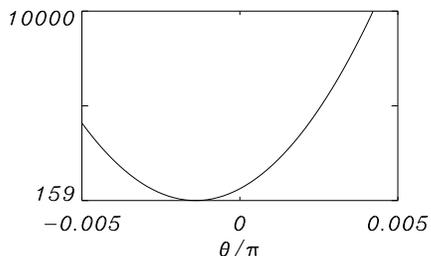}\vspace{0.3cm}}
\caption[]{Var$(\hat X_-(\theta))$
at $F=1.025$ for the critical wa\-ve\-num\-ber $k=k_c$. 
The minimum occurs for $\theta< 0$.}
\label{angle1025}
\end{figure}

A further indication of the nonlinear nature of the fluctuations
in the convective regime is given by the fact that the angle
$\theta$, for which there is the greatest reduction in the fluctuations,
changes with the strength of the pump value.  This has already been discussed 
in connection with Fig. \ref{tra.conv}($a$).  In particular, for the 
critical wave-number, $\hat X_-(\theta)$ shows strongest squeezing 
for $\theta=0$ in the linear regime below threshold. 
In the convective regime, however, the greatest reduction in the
quadrature fluctuations occurs for a value of $\theta<0$. 
This is shown in Fig. \ref{angle1025} in which we plot the 
variance  Var$(\hat X_-(\theta))$ in normal ordering and 
normalized to the shot noise level, for $F=1.025$.

The OPO can also exhibit strong correlations between the far
field intensities associated with opposite wave-numbers. We have 
calculated the fluctuations in the intensity difference for opposite 
wave-numbers associated with the normally ordered moment
$\langle:\left(\hat{A_1}^\dagger(k)\hat{A_1}(k)-\hat{A_1}^\dagger(-k)
\hat{A_1}(-k)\right)^2:\rangle$.
A negative va\-lue for this quantity indicates a non-classical 
effect sometimes referred to as twin beams or intensity-difference
squeezing \cite{twin}.  As in our discussion of quadrature squeezing, we find
that this quantity is only negative very near to threshold 
($F=1.001$).  Further into the convective regime we 
find that the macroscopic noise associated with the formation of
a pattern increases the noise in the intensity difference.  
For $F=1.01$ we find that the intensity-difference
squeezing has been replaced by fluctuations in excess of the shot
noise level.

In summary, we have shown that quantum effects can survive above 
threshold in the convective regime but only very near to threshold.
On increasing the pump and entering further into the convective regime,
we find that non-linear effects associated with the fluctuations tend 
to distribute part of the macroscopic fluctuations into the observables
that are squeezed nearer to threshold.  This identifies walk-off as an
effective mechanism of quantum decoherence in which the
macroscopic nonlinear fluctuations present in the convective
regime overwhelm quantum effects associated with noise reduction.


\section{Conclusion}
\label{conclusions}

We have introduced a suitable method to describe the quantum properties
of macroscopic patterns sustained by quantum fluctuations in a
degenerate optical para\-me\-tric oscillator with walk-off. These patterns
appear in the convective regime and are characterized by a broad far field
spectrum with continuous competition between several wave-numbers (thus, few mode
approximations are not adequate) and by being the result of amplified quantum 
fluctuations around a unstable reference state.
Traditional linearization techniques cannot be applied
in these situations. Instead we use a time dependent parametric
approximation in which the pump field is treated as a $c$-number variable but 
driven by the $c$-number representation of the quantum sub-harmonic signal 
field. The key point is that this includes the effects of the fluctuations
in the signal on the pump which in turn act back on the signal.

Using this method we have described the quantum fluctuations in type-I OPO 
with walk-off in three regimes: below the threshold of instability, in the 
convective unstable regime and in the absolute unstable regime.

Below threshold we find that the Wigner representation has a Gaussian shape 
centered at the origin. This is the result previously found in a OPO without 
walk-off \cite{OPOale} from a linearized analysis. We also find that the 
walk-off does not destroy the existence of squeezing in suitable 
quadratures.

In the convective regime the macroscopic character
 of the fluctuations is reflected
in a extremely broad Wigner distribution where the probability 
is still centered
at the origin but the nonlinear effects lead to the  appearance of
wings in the distribution which is no longer a Gaussian. These wings are  in
fact precursors of the pair of peaks appearing in the absolutely convective 
regime. We show that squeezing in the appropriate observables can be also 
obtained in this regime but only just above threshold. The walk-off and the 
nonlinearities act as quantum decoherence mechanism, distributing part of the 
macroscopic fluctuations into the observables that were squeezed below 
threshold. Another nonlinear effect appears in the selection of the quadrature displaying
reduced fluctuations, that is no longer the one determined linearly.

In the absolutely unstable regime there are also clear indications  of
nonlinear properties associated with quantum fluctuations. 
The interplay between
walk-off and non-linearity results in a complex dynamics in which
the frequencies of
far-field modes are not  constant, giving a complicated variation of the phases.
Also, the most intense mode is not the critical mode. We find that while the
Wigner distribution for the less intense modes can be approximated by a
Gaussian (displaced from the origin and orbiting about it) this is not the case
for  the most intense modes for which the distribution of fluctuations is
asymmetric around the mean amplitude with a sharp decay at some maximum
amplitude.

Finally, our method can be used in other situations and 
systems in which there 
are large fluctuations of the signal that cannot be described by 
approximations based on linearization. This includes situations in
which the critical 
fluctuations appear at threshold for pattern generation.

\acknowledgements

This work was supported by the Quantum Structures Network of the
EU TMR programme (Project No. ERB FMRX-CT96-0077). 
RZ, MSM and PC acknowledge financial support from MCyT (Spain)
projects PB97-0141-C02-02 and BFM 2001-1108.
 SMB thanks
the Royal Society of Edinburgh and the Scottish Executive Education
and Lifelong Learning Department for financial support.

\end{multicols}

\begin{thebibliography}{99}
\bibitem{spectrum} This spectrum is sharply narrowed on passing into
the absolutely unstable regime \cite{marcoPRE}.
\bibitem{supergaussian} The expression of  supergaussian pump in all simulations is
$E_0(x)=e^{-\frac{1}{2}(x/317)^{10}}$,
with $x$ varying in the interval $[-453;453]$.   This produces a profile that 
is everywhere smooth but has a flat central plateau.
\bibitem{symm} In presence of a flat pump, equations for 
$\alpha_1(k,t)$ are invariant under the change 
$\alpha_1(k,t)\rightarrow e^{i\phi}\alpha_1(k,t)$, so that it is
equivalent to construct the $W$ distribution considering 
trajectories $\alpha_1(k,t)$ or $\alpha_1'(k,t)$. 
For non flat pump this symmetry property is generally lost. 
In the pictures we represent the $W$ distribution for the amplitudes $\alpha_1$.
\end{thebibliography}
\end{document}